\pdfoutput=1
\documentclass[11pt]{article}

\usepackage[final]{acl}

\usepackage{times}
\usepackage{latexsym}

\usepackage[T1]{fontenc}
\usepackage[utf8]{inputenc}

\usepackage{microtype}

\usepackage{inconsolata}

\usepackage{adjustbox}
\usepackage{booktabs}
\usepackage{dblfloatfix}
\usepackage{graphicx}
\usepackage{multirow}
\usepackage{multicol}
\usepackage{makecell}
\usepackage{wrapfig}
\usepackage[caption=false]{subfig}
\usepackage{amsmath}

\usepackage{cleveref}

\graphicspath{{../figures/}}

\title{Enhancing Talent Search Ranking with Role-Aware Expert Mixtures and LLM-based Fine-Grained Job Descriptions}

\author{
\\
 \textbf{Jihang Li\textsuperscript{1,}\thanks{~Equal Contribution}},
 \textbf{Bing Xu\textsuperscript{2,$*$}},
 \textbf{Zulong Chen\textsuperscript{2,$*$}},
 \textbf{Chuanfei Xu\textsuperscript{5,\dag}},
\\
 \textbf{Minping Chen\textsuperscript{1,}\thanks{~Corresponding Authors}},
 \textbf{Suyu Liu\textsuperscript{2}},
 \textbf{Ying Zhou\textsuperscript{4}},
 \textbf{Zeyi Wen\textsuperscript{1,3}},
\\
 \textsuperscript{1}HKUST (GZ), Guangzhou, China \quad
 \textsuperscript{2}Alibaba Group, Hangzhou, China \\
 \textsuperscript{3}HKUST, Hong Kong, China \quad
\textsuperscript{4} Zhijiang Lab, Hangzhou, China \\
\textsuperscript{5}Guangdong Laboratory of Artificial Intelligence and Digital Economy (SZ), Shenzhen, China 
\\
 \small{
   \textbf{Correspondence:}~xuchuanfei@gml.ac.cn, ~mchen779@connect.hkust-gz.edu.cn
 }
 \\
}

\begin{document}
\maketitle

\begin{abstract}

  Talent search is a cornerstone of modern recruitment systems, yet existing
  approaches often struggle to capture nuanced job-specific preferences, model
  recruiter behavior at a fine-grained level, and mitigate noise from
  subjective human judgments. We present a novel framework that enhances talent
  search effectiveness and delivers substantial business value through two key
  innovations: (i) leveraging LLMs to extract fine-grained recruitment signals
  from job descriptions and historical hiring data, and (ii) employing a
  role-aware multi-gate MoE network to capture behavioral differences across
  recruiter roles. To further reduce noise, we introduce a multi-task learning
  module that jointly optimizes click-through rate (CTR), conversion rate
  (CVR), and resume matching relevance. Experiments on real-world recruitment
  data and online A/B testing show relative AUC gains of 1.70\% (CTR) and
  5.97\% (CVR), and a 17.29\% lift in click-through conversion rate. These
  improvements reduce dependence on external sourcing channels, enabling an
  estimated annual cost saving of millions of CNY.
\end{abstract}

\section{Introduction}
\label{sec:introduction}
% The rapid progress of Internet technology has revolutionized online
% recruitment, making it a popular tool for jobseekers or recruiters that
% effectively matches qualified candidates with appropriate
% positions~\cite{kenthapadi2017personalized,geyik2018talent}. Talent Search is a
% pivotal component of online recruitment systems, designed to empower recruiters
% to efficiently identify the ideal candidates for specific job positions.
% \Cref{fig:recruitment} shows the recruitment process of our online
% system, where talent search plays an important role.

The rise of online recruitment platforms has revolutionized how employers and
jobseekers connect, enabling efficient matching between talents and open
positions~\cite{kenthapadi2017personalized,geyik2018talent}. A core component
of these systems is talent search, which allows recruiters to identify
qualified talents for specific job postings.

\begin{figure}[t]
  \centering
  \includegraphics[width=\columnwidth]{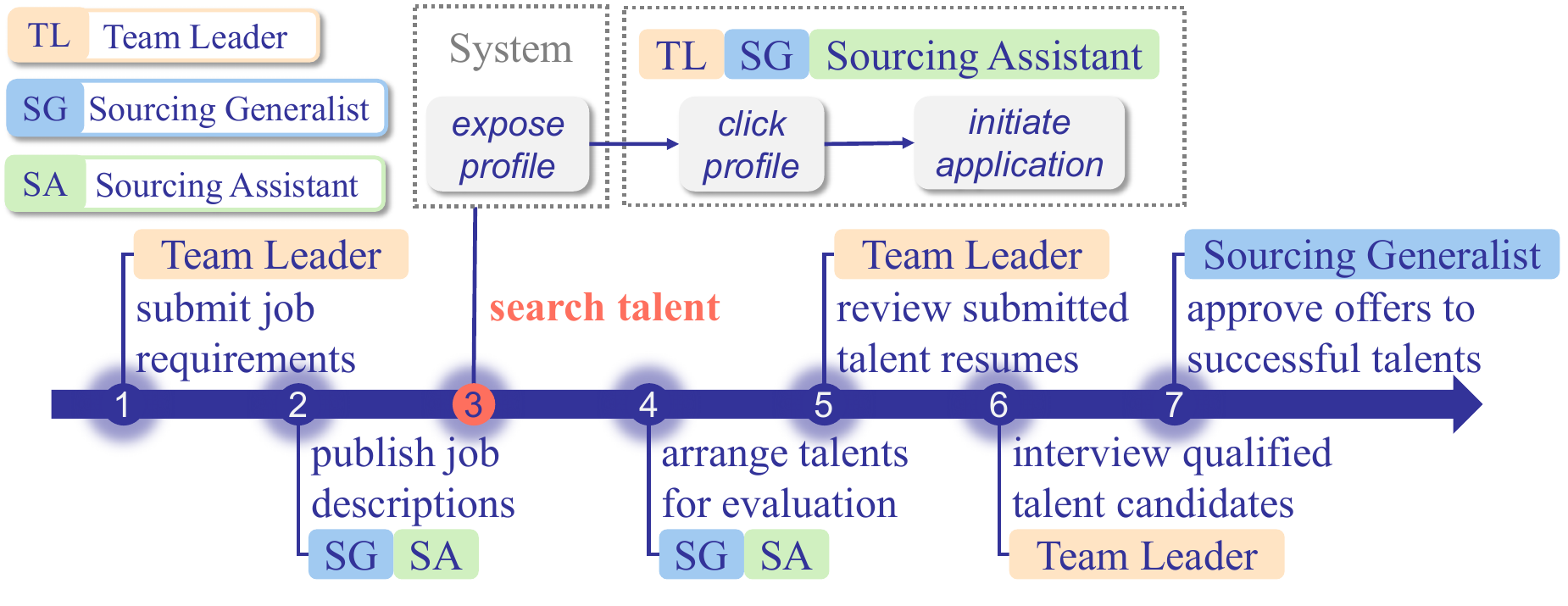}
  \caption{Recruitment process of our online system.}
  \label{fig:recruitment}
\end{figure}

As shown in \Cref{fig:recruitment}, talent search plays a central role in the
recruitment process, where recruiters interact with talent profiles by issuing
queries and browsing retrieved results. This stage involves three key steps:
(i) \textit{exposure}, where talent profiles are surfaced to the recruiter;
(ii) \textit{click}, where the recruiter views a specific profile; and (iii)
\textit{application initiation}, where the recruiter initiates a hiring
evaluation for the candidate. These interactions generate behavioral signals
such as click-through rate (CTR) and conversion rate (CVR), which are central
to assessing and optimizing search effectiveness. A detailed workflow of our
deployed talent search system is provided in \Cref{sec:system_workflow}.

% In recent years, the talent search task has focused on identifying suitable
% candidates through advanced text processing and ranking methods. Skill
% extraction from both search queries and resumes is one approach, with
% candidates ranked by skill proficiency scores~\cite{manad2018enhancing}. Deep
% learning methods, such as learning to rank using DNN
% models~\cite{ramanath2018towards} and BERT~\cite{bert} for extracting
% competency keywords~\cite{wang2021analysing}, have enhanced the candidate
% scoring performance. In addition, incorporating historical search data and
% personalization information has been shown to further improve talent search
% performance~\cite{geyik2018session,ozcaglar2019entity,yang2021cascaded}.

% Recent advances in talent search have focused on improving talent ranking
% through sophisticated text processing and modeling techniques. Approaches
% include extracting skill-related information from both recruiter queries and
% talent resumes, and ranking talents based on skill
% proficiency~\cite{manad2018enhancing}. Deep learning methods, such as learning
% to rank using deep learning networks (DNNs)~\cite{ramanath2018towards} and
% BERT-based keyword extraction~\cite{bert, wang2021analysing}, have improved
% talent scoring accuracy. Incorporating historical recruiter behavior and talent
% features has further boosted talent search
% performance~\cite{geyik2018session,ozcaglar2019entity,yang2021cascaded}.
Recent advancements in talent search have improved ranking quality through
better text modeling and personalization. Those techniques include extracting
skill signals from recruiter queries and resumes~\cite{manad2018enhancing},
using deep models for learning-to-rank~\cite{ramanath2018towards}, and applying
BERT-based keyword extraction~\cite{bert,wang2021analysing}. Incorporating
historical recruiter behavior and talent features has further enhanced
performance~\cite{geyik2018session,ozcaglar2019entity,yang2021cascaded}.

% While existing methods have demonstrated commendable outcomes in talent search,
% they continue to encounter specific challenges: (i) These methods often fall
% short in adequately modeling recruitment preferences tailored to a particular
% job. However, precisely identifying recruitment preferences can enhance the
% ability to select the most suitable candidates. (ii) Existing methods typically
% incorporate individual user preferences through entity-level or document-level
% encoding. However, this strategy can be substantially refined by utilizing more
% fine-grained information and advanced encoding techniques. (iii) Variations in
% professional expertise and subjective judgment among personnel introduce a
% degree of noise in talent search clicks and job applications, thereby impacting
% the overall effectiveness of the talent search process.

Despite these advancements, several challenges remain. (i) \textit{Insufficient
modeling of recruitment preferences}: Existing methods often fail to accurately
capture the nuanced requirements of individual job postings, limiting the
relevance of retrieved talents. (ii) \textit{Lack of role-aware
personalization}: Existing methods typically treat all recruiters uniformly,
ignoring role-specific behavioral patterns. This omission limits
personalization and leads to mismatches between recruiter intent and retrieved
talents. As shown in Table~\ref{tab:role_diff}, sourcing assistants (SA) generate nearly 50\% of page views, yet their AUC and resume pass rates are much lower than those of sourcing generalists (SG) and team leaders (TL).  (iii) \textit{Noisy behavioral signals}: Subjective judgments and
recruiter expertise gaps introduce role-dependent noise into interaction data.
Without accounting for this, models struggle to learn reliable predictors,
reducing matching accuracy.

\begin{table}[t]
  \centering
  \caption{Differences in page view (PV) rates, AUC, and resume evaluation pass
  rates across various recruiter roles. Due to confidentiality reasons, we
  cannot show the actual values of the pass rate of resume evaluation.}
  \label{tab:role_diff}

  \begin{adjustbox}{max width=\columnwidth}
  % \newcommand{\tabincell}[2]{\begin{tabular}{@{}#1@{}}#2\end{tabular}}

  % \resizebox{\columnwidth}{%
  \begin{tabular}{lccc}
    % \toprule
    \Xhline{2\arrayrulewidth}
    Role & PV Rate &   AUC & Pass Rate \\
    % \midrule
    \hline
      SA & 49.40\% & 0.554 & 2/3P      \\
      SG & 11.66\% & 0.694 & P         \\
      TL & 38.91\% & 0.693 & P         \\
    % \bottomrule
    \Xhline{2\arrayrulewidth}
  \end{tabular}
  % }
  \end{adjustbox}
\end{table}

\begin{figure*}[t]
  \centering
  \includegraphics[width=\textwidth]{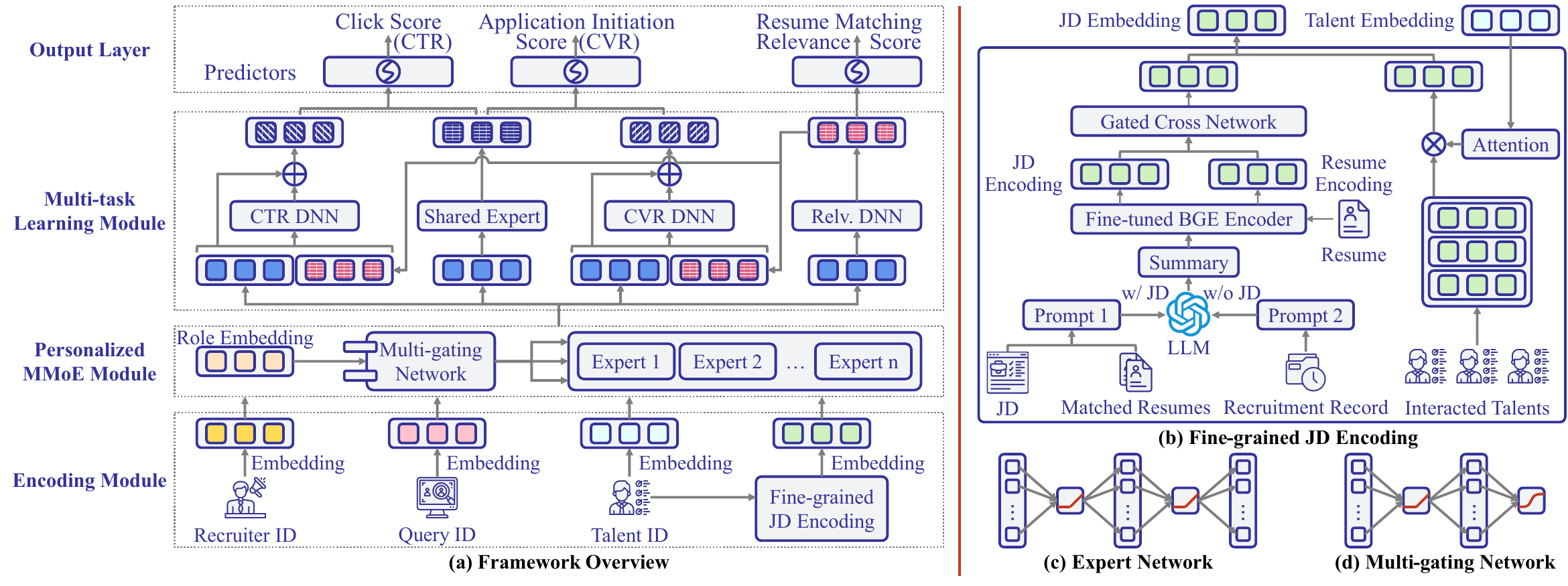}
  \caption{Overview of proposed framework.}
  \label{fig:model}
\end{figure*}

% To address these issues, we propose a novel framework that enhances talent
% search along three dimensions. First, we leverage large language models (LLMs)
% to extract fine-grained recruitment preferences by summarizing job descriptions
% and analyzing historical hiring data, using chain-of-thought (CoT)
% prompting~\cite{Wei0SBIXCLZ22}. Second, we introduce a role-aware
% multi-gate mixture-of-experts (MMoE) network to capture recruiter behavioral
% differences across roles and personalize talent ranking. Third, we incorporate
% a multi-task learning module that jointly models CTR, CVR, and resume matching
% relevance to calibrate noisy recruiter behaviors and improve prediction
% robustness.
To address these challenges, we propose a novel framework that enhances
talent search along three dimensions. First, we use large language models
(LLMs) with chain-of-thought (CoT) prompting~\cite{Wei0SBIXCLZ22} to extract
fine-grained recruitment preferences from job descriptions and historical
hiring records. Second, we design a role-aware multi-gate mixture-of-experts
(MMoE) network to model behavioral differences across recruiter roles and
personalize talent ranking. Third, we introduce a multi-task learning module
that jointly models CTR, CVR, and resume relevance to mitigate behavioral noise
and improve prediction robustness.

Our main contributions are as follows:

% \begin{enumerate}
%   \item We introduce a novel LLM-guided job representation method that extracts
%     fine-grained recruitment preferences by jointly analyzing job descriptions
%     and historical hiring records. This enables the model to capture implicit
%     signals beyond surface-level keywords, significantly improving job-talent
%     alignment.
%   \item We design a role-aware MMoE architecture that explicitly models
%     recruiter behavioral heterogeneity across different organizational roles.
%     By leveraging role-specific gating and expert routing, our model adapts
%     ranking strategies to recruiter intent with high fidelity.
%   \item We validate our framework through extensive offline experiments and
%     online A/B testing on a production-scale platform. Our method achieves
%     relative AUC gains of 1.70\% (CTR) and 5.97\% (CVR), and boosts
%     click-through conversion rate (CTCVR) by 17.29\%. These translate into an
%     estimated cost reduction of over ten million CNY annually by reducing
%     reliance on third-party sourcing channels.
% \end{enumerate}
\begin{enumerate}
  \item We introduce a novel LLM-guided job representation framework that
    extracts fine-grained recruitment preferences by jointly analyzing job
    descriptions and historical hiring records. This enables capturing implicit
    signals beyond surface-level keywords, significantly improving job-talent
    alignment.
  \item We design a role-aware MMoE architecture that models recruiter
    behavioral heterogeneity across different organizational roles. By
    leveraging role-specific gating and expert routing, our framework adapts
    ranking strategies to recruiter intent with high fidelity.
  \item We validate our framework through extensive offline experiments and
    online A/B testing on a production-scale platform, achieving relative AUC
    gains of 1.70\% (CTR) and 5.97\% (CVR), and boosting click-through conversion
    rate (CTCVR) by 17.29\%. These improvements translate into an estimated
    cost reduction of millions of CNY annually.
\end{enumerate}

% \begin{figure*}[t]
%   \centering
%   \includegraphics[width=\textwidth]{model_v3.pdf}
%   \caption{Overview of proposed method.}
%   \label{fig:model}
% \end{figure*}

\section{Methodology}
% In this section, we first introduce the problem statement and the overview of our method, and then we elaborate on the technical details of our method.

This section presents our framework to enhancing talent search via personalized
modeling and preference-aware ranking, starting with the problem statement and
followed by method details.

\subsection{Problem Statement}
\label{sec:problem_state}
Given a recruiter-issued search query or job description (JD), talent search
can be viewed as an information retrieval task: retrieving a relevant subset of
talents and ranking them by match quality. To personalize this process, we
incorporate recruiter-specific information such as recruiter ID and role type
to capture behavioral differences.

Our framework outputs three scores for each profile: (i) click-through rate (CTR) that measures the
likelihood that an exposed profile is clicked; (ii) conversion rate (CVR) that measures the
likelihood that a clicked profile leads to application initiation; and (iii) resume matching relevance. The final ranking score is computed as the product of CTR and CVR, capturing both recruiter engagement and candidate fit. This aligns with our business-to-business (B2B) platform's objectives, where CTR and CVR are key performance metrics. Therefore, we treat CTR and CVR prediction as primary tasks, and resume relevance as an auxiliary task to enhance the overall quality.

\subsection{Method Overview}
% The overview of our model is illustrated in Figure~\ref{fig:model}. The
% encoding module is responsible for converting the inputs, including the user
% ID, query ID, talent ID and job description (JD) into embeddings. One of the
% important goals of our method is to effectively capture the essential
% recruitment preferences associated with the target job. To accomplish this, we
% introduce a fine-grained job description encoding module, as shown in the right
% part of Figure~\ref{fig:model}. This module utilizes data from historically
% matched talents, historical recruitment information, interaction sequences, and
% talent resumes to model critical recruitment preferences. After obtaining the
% embeddings, they are fed into the personalized MMoE module, aiming to learn
% behavioral and preference patterns across different user roles by leveraging
% the user role information. Moreover, variations in professional expertise and
% subjective judgments introduce considerable noise in talent search clicks and
% job applications, thereby affecting the precision of CTR/CVR prediction. To
% address this issue, we propose to calibrate CTR/CVR prediction through
% multi-task learning based on the resume evaluation information.

An overview of our framework architecture is shown in \Cref{fig:model}, which
consists of three major components: (i) an \textbf{encoding module} that
transforms the inputs into embeddings, (ii) a \textbf{personalized MMoE module}
that models diverse recruiter behaviors and preferences by leveraging
role-specific embeddings, and (iii) a \textbf{multi-task learning module} that
jointly predicts click score, application initiate score, and resume matching
relevance score to calibrate predictions and reduce behavioral noise. To model
job-specific recruitment preferences, we introduce a \textbf{fine-grained JD
encoding} component that leverages LLMs and historical recruitment data. The
outputs of all modules are integrated to produce ranking scores tailored to
recruiter intent and talent suitability.

\subsection{Encoding Module}
We construct a vocabulary of recruiter, query, and talent IDs. As most queries
are short (e.g., single keywords), we encode only the query ID rather than raw
text. Embedding matrices are randomly initialized and trained end-to-end, with
recruiter, query, and talent embeddings retrieved via ID lookups.

% To capture nuanced recruitment preferences, we incorporate a fine-grained JD
% encoding module, illustrated on the right side of \Cref{fig:model}. Job
% postings often imply implicit preferences---for example, a backend developer
% position may favor talents with a computer science background. To extract such
% preferences, we enrich the JD representation using several types of historical
% data. First, we use an LLM to summarize recruitment preferences based on the JD
% text and resumes of historical matches. If the JD is unavailable as a query, we
% instead generate a talent profile using historical recruitment data, such as
% prior candidates' education and work experience. Two CoT prompts guide the LLM
% in this process. A detailed prompt example is provided in \Cref{sec:prompt}.
% The LLM-generated content is encoded using a fine-tuned BGE
% model~\cite{chen2024bge}. In parallel, the resume of the current talent is also
% encoded. We then apply a Gated Cross Network (GCN)~\cite{wang2023towards} to
% model interactions between the JD and the talent resume:
To capture nuanced recruitment preferences, we introduce a fine-grained JD
encoding module (right side of \Cref{fig:model}). Job postings often imply
implicit preferences---e.g., a backend role may favor candidates with a
computer science background. To extract such signals, we use an LLM to
summarize preferences from JD text and resumes of historical matches. If the JD
is unavailable, we instead synthesize a candidate profile from historical data
(e.g., education and work experience). Two CoT prompts guide this
process (see \Cref{sec:prompt}). The generated summaries are encoded using a
fine-tuned BGE model~\cite{chen2024bge}, alongside the current talent's resume.
We apply a Gated Cross Network (GCN)~\cite{wang2023towards} to model
interactions between JD and resume embeddings:
\begin{align*}
  c = c_0 \odot (W^{(c)} \times c_0 + b) \odot \sigma(W^{(g)} \times c_0) + c_0,
\end{align*}
where $c_0$ is the concatenation of the JD and resume embeddings produced by
the BGE encoder, $W^{(c)}$ (cross matrix) and $W^{(g)}$ (gate matrix) are
learnable weight matrices, $b$ is a bias term, and $\sigma(\cdot)$ is the
activation function.

% To further integrate recruiter's behavior history, we apply a multi-head
% attention mechanism~\cite{vaswani2017attention} between the current and
% previously interacted talents. Given a sequence of embeddings for historically
% engaged talents and the current talent embedding $e^{(t)}$, we use the sequence
% as the \textit{Key} and \textit{Value} and $e^{(t)}$ as the \textit{Query}. The
% final JD embedding is then obtained by concatenating $c$ and the attention
% output.
To incorporate recruiter behavior history, we apply multi-head
attention~\cite{vaswani2017attention} between the current talent and previously
interacted ones. Using the history as \textit{Key}/\textit{Value} and the
current embedding $e^{(t)}$ as \textit{Query}, the final JD embedding is formed
by concatenating the GCN output $c$ with the attention result.

% \begin{table}[t]
%   \centering
%   \caption{Differences in page view (PV) rates, AUC, and resume evaluation pass
%     rates across various user roles. Due to confidentiality reasons, we cannot
%     show the actual values of the pass rate of resume evaluation.}
%   \label{tab:user_role_diff}
%
%   \begin{adjustbox}{max width=\columnwidth}
%   \newcommand{\tabincell}[2]{\begin{tabular}{@{}#1@{}}#2\end{tabular}}
%   \begin{tabular}{lccc}
%   \Xhline{3\arrayrulewidth}
%   User role & PV rate &  AUC & Pass rate of resume  \\
%   \hline
%   SA & 49.40\%  & 0.554 & 2/3P \\
%   SG & 11.66\% & 0.694  & P\\
%   TL  & 38.91\% & 0.693 & P \\
%
%   \Xhline{3\arrayrulewidth}
%   \end{tabular}
%   \end{adjustbox}
% \end{table}

\subsection{Personalized MMoE Module}

Our talent search platform serves multiple recruiter roles, including sourcing
assistants (SA), sourcing generalists (SG), and team leaders (TL), each
exhibiting distinct behavioral patterns. For example in \Cref{tab:role_diff},
SA users account for nearly 50\% of page views, yet their AUC and resume pass
rates are significantly lower than those of SG and TL.

To effectively model this role-based heterogeneity, we adopt an MMoE
module~\cite{ma2018modeling}, which enables dynamic feature routing based on
recruiter role. Specifically, given an input vector $x$ (\Cref{eqn:input})
formed by concatenating the recruiter embedding $e^{(r)}$, query embedding
$e^{(q)}$, candidate embedding $e^{(t)}$, and JD embedding $e^{(j)}$, the model
processes $x$ through multiple expert networks $f_{i}$. Each expert consists of
a three-layer feedforward network with ReLU activations as illustrated in
\Cref{fig:model}c.

A role-aware gating network, as illustrated in \Cref{fig:model}d, takes the
recruiter role embedding as input and produces a softmax distribution
$\{g_{1}, g_{2}, \dots, g_{n}\}$, where $g_{i}$ represents the weight for
expert $f_{i}$ and $\sum_{i} g_{i} = 1$. The final representation $\hat{x}$ is
computed as a weighted sum of expert outputs (\Cref{eqn:mmoe}).
\begin{align}
  x &= [e^{(r)}; e^{(q)}; e^{(t)}; e^{(j)}] \label{eqn:input} \\
  \hat{x} &= \sum^{n}_{i = 1} g_{i} f_{i}(x) \label{eqn:mmoe}
\end{align}

% The outputs of these experts are aggregated using a
% role-aware gating network $g$ that takes the recruiter role embedding
% $e^{(\rho)}$ as input. Each gating network consists of a two-layer feedforward
% network with ReLU and softmax activations as depicted in \Cref{fig:model}d.
% The gating network produces a softmax distribution $\{g_{1}, \dots, g_{n}\}$,
% where $g_{i}$ represents the weight assigned to the $i$-th expert $f_{i}$ and
% $\sum^{n}_{i} g_{i} = 1$. The final representation $\hat{x}$ is a weighted sum
% of expert outputs as formulated by \Cref{eqn:mmoe}.

As outlined in \Cref{sec:problem_state}, our framework predicts three outputs. To
support this, we deploy three independent gating networks, each feeding into a
task-specific tower. Additionally, a fourth gating network is introduced to
learn shared representations across tasks, enabling the model to capture
inter-task dependencies. In total, the MMoE module maintains four gating
networks tailored for both task separation and shared behavior modeling.

% As described in \Cref{sec:problem_state}, we have three prediction tasks in the
% next multi-task learning module. Therefore, we employ three independent gating
% networks, each feeds its output to a task network as input. Additionally, to
% capture the relationships between these tasks, we employ an additional gating
% network to learn shared features across tasks. Thus, our MMoE module has four
% independent gating networks specifically.

\subsection{Multi-task Learning Module}

In addition to modeling role-specific behaviors, we also account for noise in
click and application initiation behaviors due to the variations in expertise
and subjective judgment. This behavioral noise can reduce the accuracy of CTR
and CVR predictions.

To address this, we adopt a multi-task learning approach to jointly learn CTR,
CVR, and resume matching relevance. The first two are primary prediction tasks,
while the third serves as an auxiliary task to improve overall task quality.
All tasks are formulated as binary classification. Inspired by
STEM~\cite{su2024stem}, we introduce a shared expert to capture common features
and correlations between CTR and CVR tasks, whose output is denoted as
$o_{\text{s}}$. Each task $k \in \{\text{ctr}, \text{cvr}, \text{relv}\}$ also
has a dedicated DNN tower $h_{k}$ for learning task-specific features. To
enhance task interactions, we inject the resume matching relevance output into
the input of the CTR and CVR towers:
\begin{align*}
  o_{\text{ctr}} &= [\hat{x}_{\text{ctr}}; o_{\text{relv}}] + h_{\text{ctr}}([\hat{x}_{\text{ctr}}; t_{\text{relv}}]), \\
  o_{\text{cvr}} &= [\hat{x}_{\text{cvr}}; o_{\text{relv}}] + h_{\text{cvr}}([\hat{x}_{\text{cvr}}; t_{\text{relv}}]), \\
  o_{\text{relv}} &= h_{\text{relv}}(\hat{x}_{\text{relv}}).
\end{align*}
Final predictions are computed by applying a linear projection $l(\cdot)$ followed by
softmax:
\begin{align*}
  y_{\text{ctr}} &= \text{softmax}(l([o_{\text{ctr}}; o_{\text{s}}])), \\
  y_{\text{cvr}} &= \text{softmax}(l([o_{\text{cvr}}; o_{\text{s}}])), \\
  y_{\text{relv}} &= \text{softmax}(l(o_{\text{relv}})).
\end{align*}
During training, we minimize the total loss defined as a weighted sum of
cross-entropy losses $\mathcal{L} = \sum_{k} \lambda_{k} \mathcal{L}_{k}$,
where $\lambda_{k}$ is the loss weight and $\mathcal{L}_{k}$ is the loss for
task $k$. This design allows the auxiliary task to directly inform the primary
objectives while enabling robust and noise-tolerant learning.

\section{Experiments}

\subsection{Experiment Settings}
% \textbf{Data.} There is no public talent search dataset suitable for our
% scenario, which involves job descriptions as long queries in addition to the
% general keyword-based short queries. Therefore, we collect the experimental
% data from our online recruitment system. Our training and testing sets are
% derived from user search clicks and application interactions within our
% recruitment system. The training set comprises approximately two months of
% real-world interaction data, encompassing around 495,000 samples from roughly
% 1,000 users, 200,000 talents, and 500 job positions. The testing set includes
% nearly three weeks of real-world online data, consisting of about 257,000
% samples.
\paragraph{Datasets} To the best of our knowledge, no public dataset matches
the characteristics of our talent search scenario, which involves both short
keyword queries and long job descriptions. As a result, we collect data from
our proprietary online recruitment system. The training set spans approximately
two months of interaction logs, including search clicks and application
actions. It contains around 495{,}000 samples from roughly 1{,}000 recruiters,
covering 200{,}000 talent profiles and 500 job postings. The test set comprises
about 257{,}000 samples collected over a subsequent three-week period.

% \textbf{Baselines.} Our baselines consist of a single-task learning model and
% several multi-task learning models for recommendation systems.
% GDCN~\cite{wang2023towards} is a single-task learning model which captures
% high-order feature interactions and uses an information gate to filter
% significant ones. We also compare with ESSM~\cite{ma2018entire},
% MMoE~\cite{ma2018modeling}, PLE~\cite{tang2020progressive} and
% STEM~\cite{su2024stem}, which are multi-task learning models and are introduced
% in Appendix~\ref{appendix_related_work}. Among these baselines, STEM is the
% current state-of-the-art multi-task learning model in recommender systems.
%
% We report several metrics for the CTR prediction task and the CVR prediction
% task, including AUC, MRR@10, and average precision (AP) as evaluation metrics
% in the experiments. For other implementation details, please refer to
% Appendix~\ref{implementation}.
\paragraph{Baselines} We compare our proposed framework against both
single-task and multi-task learning models commonly used in recommendation
systems. GDCN~\cite{wang2023towards} is a single-task model that captures
high-order feature interactions using a gated mechanism to retain informative
signals. For multi-task baselines, we include ESSM~\cite{ma2018entire},
MMoE~\cite{ma2018modeling}, PLE~\cite{tang2020progressive}, and
STEM~\cite{su2024stem}, the current state-of-the-art in multi-task
recommendation. These models are further described in
\Cref{sec:appendix_related_work}.

\begin{table*}[t]
  \centering
  \caption{Performance comparison among different methods. The best results are
    in \textbf{bold}.}
  \label{tab:pefromance}

  \begin{adjustbox}{max width=0.85\textwidth}
  \begin{tabular}{llccc|ccc}
    % \toprule
    \Xhline{2\arrayrulewidth}
                                     ~ & \multirow{2}{*}{Method} &                                     \multicolumn{3}{c|}{CTR} &                                      \multicolumn{3}{c}{CVR} \\
                                     ~ &                       ~ &                AUC &             MRR@10 &                 AP &                AUC &             MRR@10 &                 AP \\
    % \midrule
    \hline
    Single-task Learning               &                    GDCN &             0.6456 &             0.0131 &             0.2816 &             0.6389 &             0.0155 &             0.0484 \\
    % \midrule
    \hline
    \multirow{5}*{Multi-task Learning} &                    ESMM &             0.7067 & \underline{0.0134} &             0.2988 &             0.7005 & \underline{0.0163} &             0.0633 \\
                                     ~ &                    MMoE & \underline{0.7106} &             0.0130 &             0.3010 &             0.7108 &             0.0152 & \underline{0.0644} \\
                                     ~ &                     PLE &             0.6959 &             0.0132 &             0.2974 &             0.6884 &             0.0144 &             0.0624 \\
                                     ~ &                    STEM &             0.7091 &             0.0132 & \underline{0.3038} & \underline{0.7154} &             0.0152 &             0.0613 \\
    % \cmidrule{2-8}
    \cline{2-8}
                                     ~ &                    Ours &    \textbf{0.7227} &    \textbf{0.0136} &    \textbf{0.3139} &    \textbf{0.7581} &    \textbf{0.0192} &    \textbf{0.0680} \\
    % \bottomrule
    \Xhline{2\arrayrulewidth}
  \end{tabular}
  \end{adjustbox}
\end{table*}

\paragraph{Metrics} We evaluate performance on click-through rate (CTR) and conversion rate (CVR) prediction tasks using three standard metrics: AUC, mean reciprocal rank at 10 (MRR@10), and average precision (AP). For the online evaluation, we leverage the click-through conversion rate (CTCVR) metric. Definitions of CTR, CVR, and CTCVR are as follows:
\begin{align*}
  CTR &= \frac{N_{\rm clicks}}{N_{\rm impressions}}, \\
  CVR &= \frac{N_{\rm applications}}{N_{\rm clicks}}, \\
  CTCVR &= \frac{N_{\rm applications}}{N_{\rm impressions}},
\end{align*}
where $N_{\rm clicks}$ denotes the number of clicked talent resumes, $N_{\rm impressions}$ denotes the number of exposed talents, and $N_{\rm applications}$ denotes the number of clicked resumes that lead to application initiations. Intuitively, CTR measures the likelihood that an exposed talent resume is clicked and thus reflects recruiter engagement at the profile-viewing stage. CVR captures the probability that a clicked resume leads to an application initiation, indicating how well the clicked candidates meet recruiter requirements. CTCVR is an end-to-end efficiency metric, measuring the fraction of all exposed resumes that result in applications. Higher CTCVR signifies that recruiters are finding suitable candidates more quickly, reducing reliance on external sourcing channels. In our production system, CTR and CVR are treated as primary optimization objectives, while CTCVR serves as an aggregate business indicator for recruitment efficiency.

Other implementation details, including hyperparameters and
training protocols, are provided in \Cref{sec:implementation}.

\subsection{Offline Evaluation}
% We show the performance of different methods in \Cref{tab:pefromance}. It can
% be observed that our method outperforms both the single-task learning and the
% multi-task learning baselines, with an improvement of 1.2\% and 4.2\% in AUC
% for the CTR task and CVR task, respectively. For other metrics, our method
% also achieves significant improvements or comparable performance compared
% with the baselines, e.g., our method outperforms the baselines by 3.6\% and
% 2.9\% in AP and MRR@10 respectively on the CVR task. This indicates the
% effectiveness of our method. Additionally, we observe that all multi-task
% learning methods outperform the single-task learning method in most metrics,
% showing that multi-task learning is beneficial to improve the performance of
% each task.

The performance of all methods is reported in \Cref{tab:pefromance}. Our
proposed framework consistently outperforms both single-task and multi-task
learning baselines across all metrics. In terms of AUC, it achieves absolute
gains of 0.0121 (1.70\% relative) on the CTR task and 0.0427 (5.97\% relative)
on the CVR task over the best-performing baselines (underlined in the table).
For additional metrics, our framework also demonstrates strong performance. On
the CVR task, it improves MRR@10 by 0.0029 (17.79\% relative) and AP by 0.0036
(5.59\% relative) compared to the best baseline. These results highlight the
effectiveness of our approach in improving both ranking quality and predictive
accuracy. Moreover, we observe that all multi-task learning methods outperform
the single-task baseline in most metrics, confirming the benefit of multi-task
learning for jointly modeling related objectives such as CTR and CVR
prediction.

\subsection{Online A/B Test}
\begin{wrapfigure}[10]{r}{0.45\columnwidth}
  \centering
  \vspace{-0.5\baselineskip}
  \includegraphics[width=1.0\linewidth]{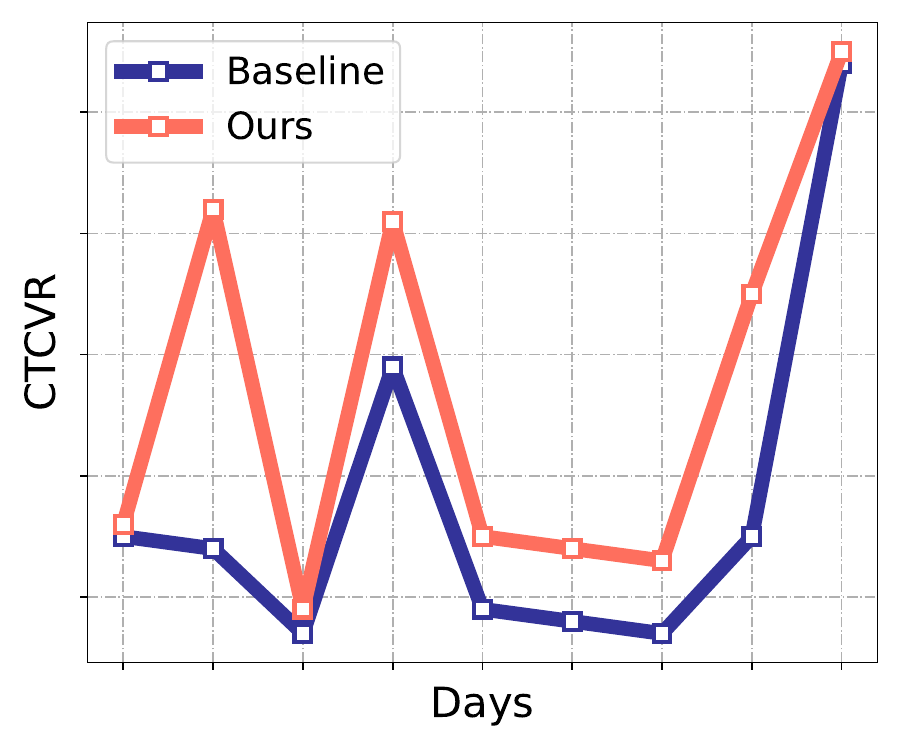}
  \caption{A/B test result. Values are concealed due to confidentiality.}
  \label{fig:ab_test}
\end{wrapfigure}

In addition to our offline evaluations, we conducted an online A/B test to
assess the real-world effectiveness of our framework. The results are shown in
\Cref{fig:ab_test}. We use GDCN as the online baseline and deploy our framework
on the personalized recommendation platform using Google's standard
experimentation protocol. User traffic was randomly split 1:1 between the experimental group (ours) and
the control group (baseline). The test ran for nine consecutive working days to
ensure stability and statistical significance. 
\setlength{\abovecaptionskip}{0.25\baselineskip} % 调整标题上方空白
Compared to the baseline, our framework achieved a 17.29\% relative improvement in click-through conversion
rate (CTCVR). Importantly, the performance gain is statistically significant,
with a $p$-value of 0.01598 ($<$ 0.05), validating the practical impact of our
approach in a production environment. Furthermore, with the assistance of our framework, we yield annual savings of millions of CNY. Specifically, nearly 40\% of hires already come from our internal talent database, which incurs no channel fees. By boosting conversions (CTCVR) from this source, we further reduce reliance on external channels, e.g., headhunting channels, thus saving significant external channel fees.
% Due to confidentiality constraints, daily CTCVR values are not
% disclosed.

\subsection{Ablation Study}
% \textbf{Ablation Study.} To verify the effectiveness of different modules in
% our method, we conduct an ablation study, as shown in the upper part of
% Table~\ref{ablation_study}. Specifically, \textit{w/o JD enc.} denotes the
% model that only uses the embedding of the job description ID, \textit{w/o JD
% enc.\& MTL} denotes the model that only uses the embedding of the job
% description ID and trains the CTR and CVR prediction tasks separately, and
% \textit{w/o JD enc. \& MTL \& P-MMoE} refers to a model variant based on the
% \textit{w/o JD enc. \& MTL} configuration. In this variant, the original MMoE
% module is employed instead of the personalized MMoE module proposed in this
% paper. We can see that removing any of the above modules results in a
% performance decline for both the CTR prediction task and CVR prediction task,
% indicating their effectiveness.
\paragraph{Impact of Model Components} To evaluate the contribution of key
components in our framework, we conduct an ablation study. Results are shown in
the upper part of \Cref{tab:ablation}, which reports AUC scores for the CTR and
CVR tasks. We define the following model variants: (i) \textit{w/o JD enc.}
replaces our fine-grained job description encoder with a simple job ID
embedding, (ii) \textit{w/o JD enc. \& MTL} further removes multi-task learning
by training CTR and CVR predictors independently, and (iii) \textit{w/o JD enc.
\& MTL \& P-MMoE} additionally replaces our personalized MMoE module with a
standard MMoE architecture. We observe that removing any of these modules leads
to a noticeable drop in AUC for both CTR and CVR tasks. This confirms the
importance of fine-grained job representation, multi-task learning, and
personalized expert routing in improving model performance.

\begin{table}[t]
  \centering
  \caption{Ablation study results (AUC score).}
  \label{tab:ablation}

  \begin{adjustbox}{max width=\columnwidth}
  \begin{tabular}{lccc}
    \Xhline{2\arrayrulewidth}
    Model                         & CTR             & CVR             & Avg.            \\
    \hline
    Ours (full model)             & \textbf{0.7227} & \textbf{0.7581} & \textbf{0.7404} \\
    w/o JD enc.                   & 0.7182          & 0.7543          & 0.7363          \\
    w/o JD enc. \& MTL            & 0.7164          & 0.7264          & 0.7214          \\
    w/o JD enc. \& MTL  \& P-MMoE & 0.6970          & 0.7069          & 0.7020          \\
    \hline
    Ours (w/~~ LLM in JD enc.)   & 0.7227          & \textbf{0.7581} & \textbf{0.7404} \\
    Ours (w/o LLM in JD enc.)     & \textbf{0.7282} & 0.7374          & 0.7328          \\
    \hline
    MMoE                          & 0.7106          & 0.7108          & 0.7107          \\
    MMoE + our LLM in JD enc.     & \textbf{0.7114} & \textbf{0.7384} & \textbf{0.7294} \\
    \Xhline{2\arrayrulewidth}
    \end{tabular}
  \end{adjustbox}
\end{table}

\paragraph{Impact of LLM-Based Recruitment Preference Summarization} We also
evaluate the effectiveness of using an LLM to summarize key recruitment
preferences for job descriptions. Specifically, we compare our framework with a
variant that removes the LLM and instead uses simple text concatenation.
Results are shown in the middle part of \Cref{tab:ablation}. Since CTR and CVR
predictions are used jointly for ranking, we report the average AUC to assess
overall ranking quality. The results indicate that incorporating LLM-based
summaries improves model performance. To further validate this, we integrate
our LLM-based job description encoder into one of the baselines (MMoE). As
shown in the lower part of \Cref{tab:ablation}, this enhanced version (MMoE +
LLM) yields consistent gains over the vanilla MMoE model on both tasks,
demonstrating the general effectiveness of using LLMs for recruitment
preference modeling.

% \textbf{Effect of the Number of Experts.} Here, we conduct experiments to
% investigate the impact of the number of experts in the personalized MMoE
% module, as shown in Figure~\ref{impact_exp}(a). Each expert is expected to
% learn the personalized features for each use role. We can see that using
% three experts achieves the best performance, which aligns with the number of
% user roles (SA, SG, TL in Table~\ref{tab:role_diff}) in our system. Besides,
% increasing the number of experts to 5 and 10 leads to a performance drop, as
% it may increase the risk of overfitting. Therefore, we use three experts in
% our model, which is intuitive and interpretable.

\begin{figure}[t]
  \centering
  \subfloat[]{
    \includegraphics[width=0.45\linewidth]{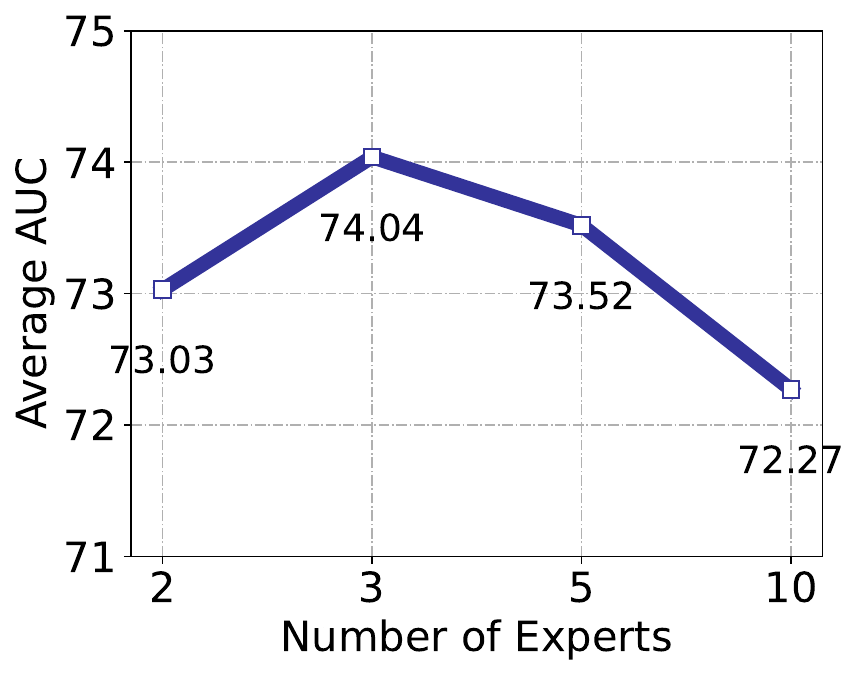}
    \label{fig:num_of_moe}
  }
  \centering
  \subfloat[]{
    \includegraphics[width=0.45\linewidth]{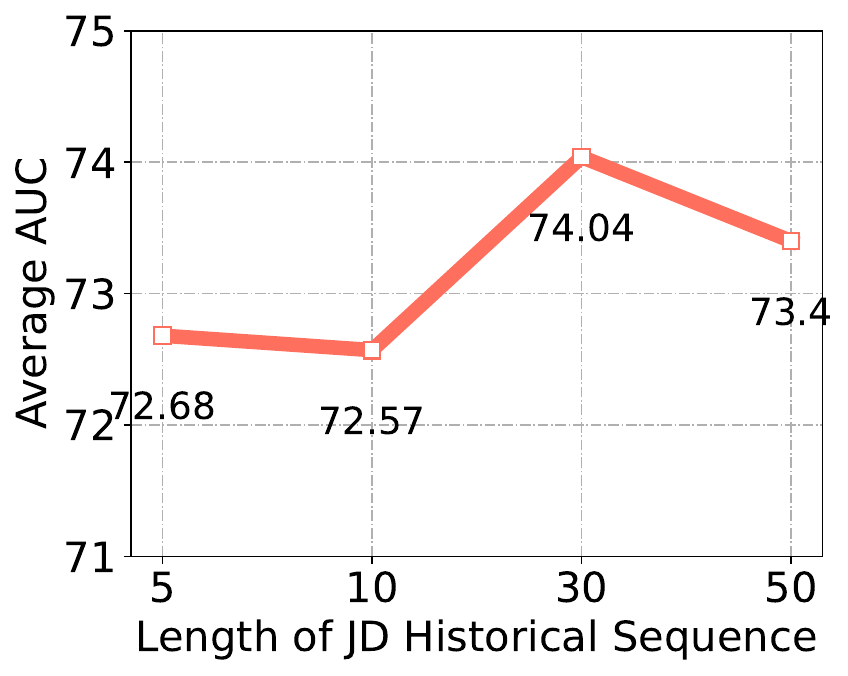}
    \label{fig:seqlen_of_interact}
  }

  \caption{Impact of the number of experts in the MMoE module and the
  historical sequence length in the JD encoding module on the final
  performance. The performance is the average AUC result of the CTR prediction
  task and the CVR prediction task.}
  \label{fig:impact_exp}
\end{figure}

\paragraph{Impact of the Number of Experts} We further analyze how the number
of experts in the personalized MMoE module affects model performance, as
illustrated in \Cref{fig:num_of_moe}. Since each expert is designed to capture
role-specific patterns, we expect optimal performance when the number of
experts aligns with the number of recruiter roles in our system (SA, SG, TL;
see \Cref{tab:role_diff}). Empirically, using three experts yields the best
performance. Increasing the number of experts to five or ten results in
decreased performance, likely due to overfitting and increased model
complexity. Based on these observations, we adopt three experts in our final
model, which provides both strong empirical results and a straightforward,
interpretable mapping to the three recruiter roles.

\begin{figure*}[t]
  \centering
  \includegraphics[width=\textwidth]{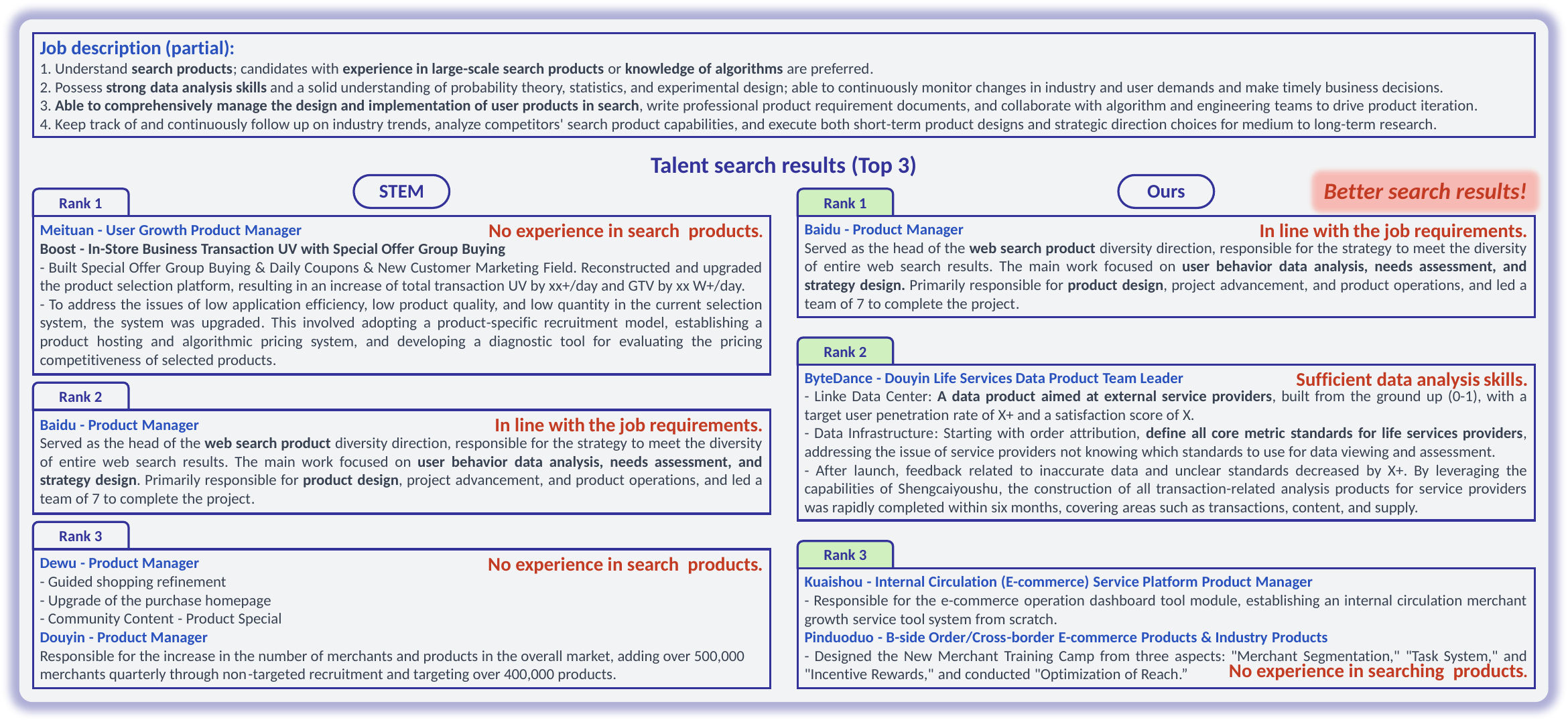}
  \caption{Case study of candidate ranking for a ``Search Product Manager'' position.}
  \label{fig:case_study}
\end{figure*}

% \textbf{Effect of the Length of Historical Sequence.} We also explore the
% impact of the length of the historical sequence used in the job description
% (JD) encoding module, and the results are shown in
% Figure~\ref{fig:impact_exp}(b). In the JD encoding module, we utilize
% historical information including the historically matched resumes and the
% historically interacted talents. We find that using a sequence length of 30
% achieves the best performance, and using sequence lengths shorter or longer
% than 30 achieves sub-optimal performance. Two reasons may lead to this
% phenomenon: (i) LLM or the model cannot learn sufficient features when little
% historical information is provided. (ii) When the length of the historical
% sequence is longer than 30, it may introduce some noise to the model, thereby
% destroying the model performance.
\paragraph{Impact of Historical Sequence Length} We examine how the length of
the historical sequence used in the JD encoding module influences model
performance. As shown in \Cref{fig:seqlen_of_interact}, this module
incorporates both historically matched resumes and previously interacted
talents. We observe that a sequence length of 30 yields the best performance.
Shorter sequences may fail to provide sufficient context for the LLM or the
downstream model to extract meaningful features. Conversely, longer sequences
(e.g., beyond 30) may introduce noise or less relevant historical data, which
can negatively impact performance. These findings suggest that using a
well-balanced sequence length is important for maximizing the utility of
historical information in JD encoding.

% \textbf{Online A/B Test.} In addition to the offline experiments, we conduct an online A/B test to demonstrate the real-world effectiveness of our method. The results are shown in Figure~\ref{ab_test_result}. In more detail, we use GDCN as the online baseline, and implement the A/B test on our personalized recommendation platform using Google's experimental method.
% \setlength{\abovecaptionskip}{1.2pt} % 调整标题上方空白
% \begin{wrapfigure}[9]{r}{0.2\textwidth}
%     \centering
%     \includegraphics[width=1.0\linewidth]{ab_test.png}
%     \caption{Results of online A/B test.}
%     \label{ab_test_result}
% \end{wrapfigure}
% Traffic was randomly allocated in a 1:1 ratio between the experimental and control groups, and the test was observed online for nine working days to ensure data stability. Our method achieves a relative improvement of 17.29\% in CTCVR during the nine days compared with the baseline. Due to confidentiality reasons, the actual CTCVR value of each day is not presented here. Notably, the performance difference between our method and the baseline is statistically significant with a $p$-value=0.01598 $<$ 0.05, indicating the effectiveness of our method for real-world applications.

\subsection{Case Study}
% Here, we present a case study which compares two sets of talent search
% results predicted by STEM and our method for a search product manager
% position, as shown in Figure~\ref{fig:case_study}. The job description
% emphasizes expertise in large-scale search products, data analysis, and
% industry trend monitoring. Among the top candidates, the Baidu Product
% Manager (Rank 1 in our method while Rank 2 in STEM) stands out as the most
% qualified, with direct experience in web search product diversity, user
% behavior analysis, and leading cross-functional teams. Notably, while STEM's
% Rank 3 candidate lacked search experience entirely, our method surfaced
% alternative specialists: a data product leader (Rank 2) with advanced
% analytical capabilities applicable to search optimization, and an e-commerce
% tools expert (Rank 3) with transferable skills in metric standardization.
% This suggests our method employs a more nuanced evaluation that balances
% direct experience (Rank 1) with adjacent competencies (Ranks 2-3), whereas
% STEM's approach overlooked critical search-specific qualifications in two of
% its top three candidates.
To qualitatively evaluate the effectiveness of our framework, we present a case
study comparing talent search results from STEM and our approach for a ``Search
Product Manager'' position. The comparison is shown in \Cref{fig:case_study}.
The job description emphasizes expertise in large-scale search systems, data
analysis, and industry trend monitoring.

% Among the top-ranked candidates, the ``Baidu -  Product Manager''---ranked 1st by our
% method and 2nd by STEM---clearly aligns with the job requirements,
% demonstrating direct experience in web search diversity, user behavior
% analysis, and cross-functional team leadership. In contrast, STEM's 3rd-ranked
% candidate lacks any relevant search experience. Our method instead surfaces two
% more contextually aligned candidates: a data product leader (Rank 2), whose
% analytical strengths are highly relevant to search optimization, and an
% e-commerce tools expert (Rank 3), whose experience in metric standardization is
% transferable to evaluating search product quality. This result suggests that
% our model provides a more nuanced ranking, effectively balancing direct domain
% experience (Rank 1) with adjacent, high-relevance competencies (Ranks 2--3),
% whereas STEM fails to capture these finer distinctions and ranks a
% less-qualified candidate among its top three.

Among the top-ranked candidates, the ``Baidu - Product Manager'' (ranked 1st by
ours and 2nd by STEM) clearly matches the job requirements, showing direct
experience in web search diversity, user behavior analysis, and team
leadership. In contrast, STEM's 3rd-ranked candidate lacks relevant search
experience. Our framework instead highlights two more contextually suitable
candidates: a data product leader (Rank 2) with strong analytical skills for
search optimization, and an e-commerce tools expert (Rank 3) with transferable
experience in metric standardization. These results suggest our framework
offers a more refined ranking by balancing domain expertise (Rank 1) with
related high-value skills (Ranks 2 and 3), while STEM overlooks such nuances
and ranks a less-qualified candidate in the top three.

\section{Related Works}
\label{sec:related_works}

Talent search aims to identify suitable candidates based on recruiter queries.
Early work, such as \citet{apatean2017machine}, explored traditional machine
learning methods like KNN~\cite{wu2008top} and LDA~\cite{blei2003latent} to
classify candidate attributes such as education, programming, and language
skills, enabling structured search over profiles. \citet{manad2018enhancing}
extended this by extracting skills from both queries and resumes and ranking
candidates by skill proficiency.

Recent methods leverage deep learning. \citet{ramanath2018towards} applied
learning-to-rank techniques using DNNs, while \citet{wang2021analysing} used
BERT~\cite{bert} to extract competency-related keywords and compute weighted
scores. Their work also introduced a Competence Map (CMAP) to model inter-skill
relationships.

Another line of research leverages recruiter interaction data to improve talent
search quality. \citet{ha2016search,ha2017query} first explored this idea using
historical search logs. \citet{geyik2018session} further incorporated real-time
recruiter feedback to infer intent clusters and applied a multi-armed bandit
framework for ranking. Similarly, \citet{ozcaglar2019entity} introduced a
two-level ranking system that integrates structured candidate features and uses
recruiter actions as supervised signals. \citet{yang2021cascaded} proposed a
cascaded architecture combining DNN and BERT, where personalized recruiter
preferences are explicitly modeled in the final ranking stage.

\section{Conclusion}
% In this paper, we introduce an innovative method for talent search that
% employs large language models (LLMs) for modeling recruitment preferences and
% a personalized multi-gate mixture-of-experts (MMoE) network to capture varied
% user behaviors. Furthermore, we incorporate a multi-task learning module to
% adjust for behavioral noise, thereby enhancing task quality. Our
% comprehensive evaluations, which include offline assessments on the
% real-world dataset and an online A/B test within our recruitment system, show
% that our method surpasses existing state-of-the-art techniques, achieving
% significant improvements of 1.2\% and 4.3\% in AUC for CTR and CVR prediction
% tasks, respectively, and a 17.29\% relative improvement in the online A/B
% test.

% We propose a novel framework for talent search that integrates LLM models to
% model recruitment preferences and a personalized MMoE network to capture
% diverse recruiter behaviors. To further enhance prediction robustness, we
% incorporate a multi-task learning module that mitigates behavioral noise across
% tasks. Extensive evaluations on real-world recruitment data demonstrate the
% effectiveness of our approach. Our method achieves significant performance
% gains over state-of-the-art baselines, with improvements of 1.2\% and 4.3\% in
% AUC for CTR and CVR prediction, respectively, and a 17.29\% relative lift in an
% online A/B test. These results highlight the potential of combining LLMs,
% role-aware personalization, and multi-task learning in real-world talent search
% systems.

We present a novel talent search framework that combines LLM-based recruitment
preference modeling, a role-aware MMoE network for capturing recruiter
heterogeneity, and a multi-task learning module to reduce behavioral noise.
Experiments on real-world data and an online A/B test demonstrate significant
relative performance gains of 1.70\% and 5.97\% in AUC for CTR and CVR, and a
17.29\% lift in click-through conversion rate (CTCVR). These improvements
translate into substantial business value, enabling an estimated annual cost
saving of millions of CNY by reducing reliance on external recruiting channels.
Our results highlight the practical impact of integrating LLMs, role-specific
modeling, and multi-task optimization in real-world talent search systems.

\section{Limitations}
Our system currently depends on sufficient historical interaction data to model
recruiter behavior and job preferences effectively. In cold-start scenarios
such as new job postings or first-time recruiters, the framework's
effectiveness may be limited. Additionally, although our framework supports
multiple recruiter roles, it does not explicitly account for temporal dynamics
in recruiter behavior or job market trends, which may evolve over time.

\section*{Acknowledgments}
This work is supported by National Key R\&D Program of China under Grant No. 2024YFA1012700, and by the Guangzhou Industrial Information and Intelligent Key Laboratory Project (No. 2024A03J0628). It is also funded by the NSFC Project (No. 62306256) and the Natural Science Foundation of Guangdong Province (No. 2025A1515010261).

\bibliography{custom}

\begin{thebibliography}{27}
\providecommand{\natexlab}[1]{#1}

\bibitem[{ope(2025)}]{opensearch}
 2025.
\newblock {OpenSearch}.
\newblock \url{https://opensearch.org}.
\newblock [Accessed 04-07-2025].

\bibitem[{Apatean et~al.(2017)Apatean, Szakacs, and Tilca}]{apatean2017machine}
Anca Apatean, Evelyn Szakacs, and Magnolia Tilca. 2017.
\newblock Machine-learning based application for staff recruiting.
\newblock \emph{Acta Technica Napocensis}, 58(4):16--21.

\bibitem[{Blei et~al.(2003)Blei, Ng, and Jordan}]{blei2003latent}
David~M Blei, Andrew~Y Ng, and Michael~I Jordan. 2003.
\newblock Latent dirichlet allocation.
\newblock \emph{Journal of machine Learning research}, 3(Jan):993--1022.

\bibitem[{Chen et~al.(2024)Chen, Xiao, Zhang, Luo, Lian, and Liu}]{chen2024bge}
Jianlv Chen, Shitao Xiao, Peitian Zhang, Kun Luo, Defu Lian, and Zheng Liu. 2024.
\newblock Bge m3-embedding: Multi-lingual, multi-functionality, multi-granularity text embeddings through self-knowledge distillation.
\newblock \emph{arXiv preprint arXiv:2402.03216}.

\bibitem[{Chen et~al.(2019)Chen, Wang, Xie, Wu, Bu, Wang, and Chen}]{chen2019co}
Zhongxia Chen, Xiting Wang, Xing Xie, Tong Wu, Guoqing Bu, Yining Wang, and Enhong Chen. 2019.
\newblock Co-attentive multi-task learning for explainable recommendation.
\newblock In \emph{IJCAI}, volume 2019, pages 2137--2143.

\bibitem[{Devlin et~al.(2019)Devlin, Chang, Lee, and Toutanova}]{bert}
Jacob Devlin, Ming{-}Wei Chang, Kenton Lee, and Kristina Toutanova. 2019.
\newblock {BERT:} pre-training of deep bidirectional transformers for language understanding.
\newblock In \emph{The Annual Conference of the North American Chapter of the Association for Computational Linguistics: Human Language Technologies (NAACL-HLT)}, pages 4171--4186.

\bibitem[{Geyik et~al.(2018{\natexlab{a}})Geyik, Dialani, Meng, and Smith}]{geyik2018session}
Sahin~Cem Geyik, Vijay Dialani, Meng Meng, and Ryan Smith. 2018{\natexlab{a}}.
\newblock In-session personalization for talent search.
\newblock In \emph{Proceedings of the 27th ACM international conference on information and knowledge management}, pages 2107--2115.

\bibitem[{Geyik et~al.(2018{\natexlab{b}})Geyik, Guo, Hu, Ozcaglar, Thakkar, Wu, and Kenthapadi}]{geyik2018talent}
Sahin~Cem Geyik, Qi~Guo, Bo~Hu, Cagri Ozcaglar, Ketan Thakkar, Xianren Wu, and Krishnaram Kenthapadi. 2018{\natexlab{b}}.
\newblock Talent search and recommendation systems at linkedin: Practical challenges and lessons learned.
\newblock In \emph{The 41st International ACM SIGIR Conference on Research \& Development in Information Retrieval}, pages 1353--1354.

\bibitem[{Ha-Thuc et~al.(2016)Ha-Thuc, Xu, Kanduri, Wu, Dialani, Yan, Gupta, and Sinha}]{ha2016search}
Viet Ha-Thuc, Ye~Xu, Satya~Pradeep Kanduri, Xianren Wu, Vijay Dialani, Yan Yan, Abhishek Gupta, and Shakti Sinha. 2016.
\newblock Search by ideal candidates: Next generation of talent search at linkedin.
\newblock In \emph{Proceedings of the 25th International Conference Companion on World Wide Web}, pages 195--198.

\bibitem[{Ha-Thuc et~al.(2017)Ha-Thuc, Yan, Wu, Dialani, Gupta, and Sinha}]{ha2017query}
Viet Ha-Thuc, Yan Yan, Xianren Wu, Vijay Dialani, Abhishek Gupta, and Shakti Sinha. 2017.
\newblock From query-by-keyword to query-by-example: Linkedin talent search approach.
\newblock In \emph{Proceedings of the 2017 ACM on Conference on Information and Knowledge Management}, pages 1737--1745.

\bibitem[{Huang et~al.(2013)Huang, He, Gao, Deng, Acero, and Heck}]{huang2013learning}
Po-Sen Huang, Xiaodong He, Jianfeng Gao, Li~Deng, Alex Acero, and Larry Heck. 2013.
\newblock Learning deep structured semantic models for web search using clickthrough data.
\newblock In \emph{Proceedings of the 22nd ACM international conference on Information \& Knowledge Management}, pages 2333--2338.

\bibitem[{Kenthapadi et~al.(2017)Kenthapadi, Le, and Venkataraman}]{kenthapadi2017personalized}
Krishnaram Kenthapadi, Benjamin Le, and Ganesh Venkataraman. 2017.
\newblock Personalized job recommendation system at linkedin: Practical challenges and lessons learned.
\newblock In \emph{Proceedings of the eleventh ACM conference on recommender systems}, pages 346--347.

\bibitem[{Ma et~al.(2018{\natexlab{a}})Ma, Zhao, Yi, Chen, Hong, and Chi}]{ma2018modeling}
Jiaqi Ma, Zhe Zhao, Xinyang Yi, Jilin Chen, Lichan Hong, and Ed~H Chi. 2018{\natexlab{a}}.
\newblock Modeling task relationships in multi-task learning with multi-gate mixture-of-experts.
\newblock In \emph{Proceedings of the 24th ACM SIGKDD international conference on knowledge discovery \& data mining}, pages 1930--1939.

\bibitem[{Ma et~al.(2018{\natexlab{b}})Ma, Zhao, Huang, Wang, Hu, Zhu, and Gai}]{ma2018entire}
Xiao Ma, Liqin Zhao, Guan Huang, Zhi Wang, Zelin Hu, Xiaoqiang Zhu, and Kun Gai. 2018{\natexlab{b}}.
\newblock Entire space multi-task model: An effective approach for estimating post-click conversion rate.
\newblock In \emph{The 41st International ACM SIGIR Conference on Research \& Development in Information Retrieval}, pages 1137--1140.

\bibitem[{Manad et~al.(2018)Manad, Bentounsi, and Darmon}]{manad2018enhancing}
Otman Manad, Mehdi Bentounsi, and Patrice Darmon. 2018.
\newblock Enhancing talent search by integrating and querying big hr data.
\newblock In \emph{2018 IEEE International Conference on Big Data (Big Data)}, pages 4095--4100. IEEE.

\bibitem[{Ozcaglar et~al.(2019)Ozcaglar, Geyik, Schmitz, Sharma, Shelkovnykov, Ma, and Buchanan}]{ozcaglar2019entity}
Cagri Ozcaglar, Sahin Geyik, Brian Schmitz, Prakhar Sharma, Alex Shelkovnykov, Yiming Ma, and Erik Buchanan. 2019.
\newblock Entity personalized talent search models with tree interaction features.
\newblock In \emph{The World Wide Web Conference}, pages 3116--3122.

\bibitem[{Ramanath et~al.(2018)Ramanath, Inan, Polatkan, Hu, Guo, Ozcaglar, Wu, Kenthapadi, and Geyik}]{ramanath2018towards}
Rohan Ramanath, Hakan Inan, Gungor Polatkan, Bo~Hu, Qi~Guo, Cagri Ozcaglar, Xianren Wu, Krishnaram Kenthapadi, and Sahin~Cem Geyik. 2018.
\newblock Towards deep and representation learning for talent search at linkedin.
\newblock In \emph{Proceedings of the 27th ACM international conference on information and knowledge management}, pages 2253--2261.

\bibitem[{Su et~al.(2024)Su, Pan, Wang, Xiao, Quan, Chen, and Jiang}]{su2024stem}
Liangcai Su, Junwei Pan, Ximei Wang, Xi~Xiao, Shijie Quan, Xihua Chen, and Jie Jiang. 2024.
\newblock Stem: unleashing the power of embeddings for multi-task recommendation.
\newblock In \emph{Proceedings of the AAAI Conference on Artificial Intelligence}, volume~38, pages 9002--9010.

\bibitem[{Tang et~al.(2020)Tang, Liu, Zhao, and Gong}]{tang2020progressive}
Hongyan Tang, Junning Liu, Ming Zhao, and Xudong Gong. 2020.
\newblock Progressive layered extraction (ple): A novel multi-task learning (mtl) model for personalized recommendations.
\newblock In \emph{Proceedings of the 14th ACM conference on recommender systems}, pages 269--278.

\bibitem[{Tay et~al.(2018)Tay, Luu, and Hui}]{tay2018multi}
Yi~Tay, Anh~Tuan Luu, and Siu~Cheung Hui. 2018.
\newblock Multi-pointer co-attention networks for recommendation.
\newblock In \emph{Proceedings of the 24th ACM SIGKDD international conference on knowledge discovery \& data mining}, pages 2309--2318.

\bibitem[{Vaswani(2017)}]{vaswani2017attention}
A~Vaswani. 2017.
\newblock Attention is all you need.
\newblock \emph{Advances in Neural Information Processing Systems}.

\bibitem[{Wang et~al.(2023)Wang, Gu, Li, Lu, Zhang, and Gu}]{wang2023towards}
Fangye Wang, Hansu Gu, Dongsheng Li, Tun Lu, Peng Zhang, and Ning Gu. 2023.
\newblock Towards deeper, lighter and interpretable cross network for ctr prediction.
\newblock In \emph{Proceedings of the 32nd ACM international conference on information and knowledge management}, pages 2523--2533.

\bibitem[{Wang et~al.(2021)Wang, Allouache, and Joubert}]{wang2021analysing}
Yan Wang, Yacine Allouache, and Christian Joubert. 2021.
\newblock Analysing cv corpus for finding suitable candidates using knowledge graph and bert.
\newblock In \emph{DBKDA 2021, the thirteenth international conference on advances in databases, knowledge, and data applications}.

\bibitem[{Wei et~al.(2022)Wei, Wang, Schuurmans, Bosma, Ichter, Xia, Chi, Le, and Zhou}]{Wei0SBIXCLZ22}
Jason Wei, Xuezhi Wang, Dale Schuurmans, Maarten Bosma, Brian Ichter, Fei Xia, Ed~H. Chi, Quoc~V. Le, and Denny Zhou. 2022.
\newblock Chain-of-thought prompting elicits reasoning in large language models.
\newblock In \emph{Annual Conference on Neural Information Processing Systems (NeurIPS)}.

\bibitem[{Wu et~al.(2008)Wu, Kumar, Ross~Quinlan, Ghosh, Yang, Motoda, McLachlan, Ng, Liu, Yu et~al.}]{wu2008top}
Xindong Wu, Vipin Kumar, J~Ross~Quinlan, Joydeep Ghosh, Qiang Yang, Hiroshi Motoda, Geoffrey~J McLachlan, Angus Ng, Bing Liu, Philip~S Yu, and 1 others. 2008.
\newblock Top 10 algorithms in data mining.
\newblock \emph{Knowledge and information systems}, 14:1--37.

\bibitem[{Yang et~al.(2021)Yang, Yan, Lad, Liu, and Guo}]{yang2021cascaded}
Zimeng Yang, Song Yan, Abhimanyu Lad, Xiaowei Liu, and Weiwei Guo. 2021.
\newblock Cascaded deep neural ranking models in linkedin people search.
\newblock In \emph{Proceedings of the 30th ACM International Conference on Information \& Knowledge Management}, pages 4312--4320.

\bibitem[{Zhao et~al.(2019)Zhao, Hong, Wei, Chen, Nath, Andrews, Kumthekar, Sathiamoorthy, Yi, and Chi}]{zhao2019recommending}
Zhe Zhao, Lichan Hong, Li~Wei, Jilin Chen, Aniruddh Nath, Shawn Andrews, Aditee Kumthekar, Maheswaran Sathiamoorthy, Xinyang Yi, and Ed~Chi. 2019.
\newblock Recommending what video to watch next: a multitask ranking system.
\newblock In \emph{Proceedings of the 13th ACM conference on recommender systems}, pages 43--51.

\end{thebibliography}

% \vfill\null

\appendix

\section{System Workflow}
\label{sec:system_workflow}
\begin{figure}[htb]
  \centering
  \includegraphics[width=\columnwidth]{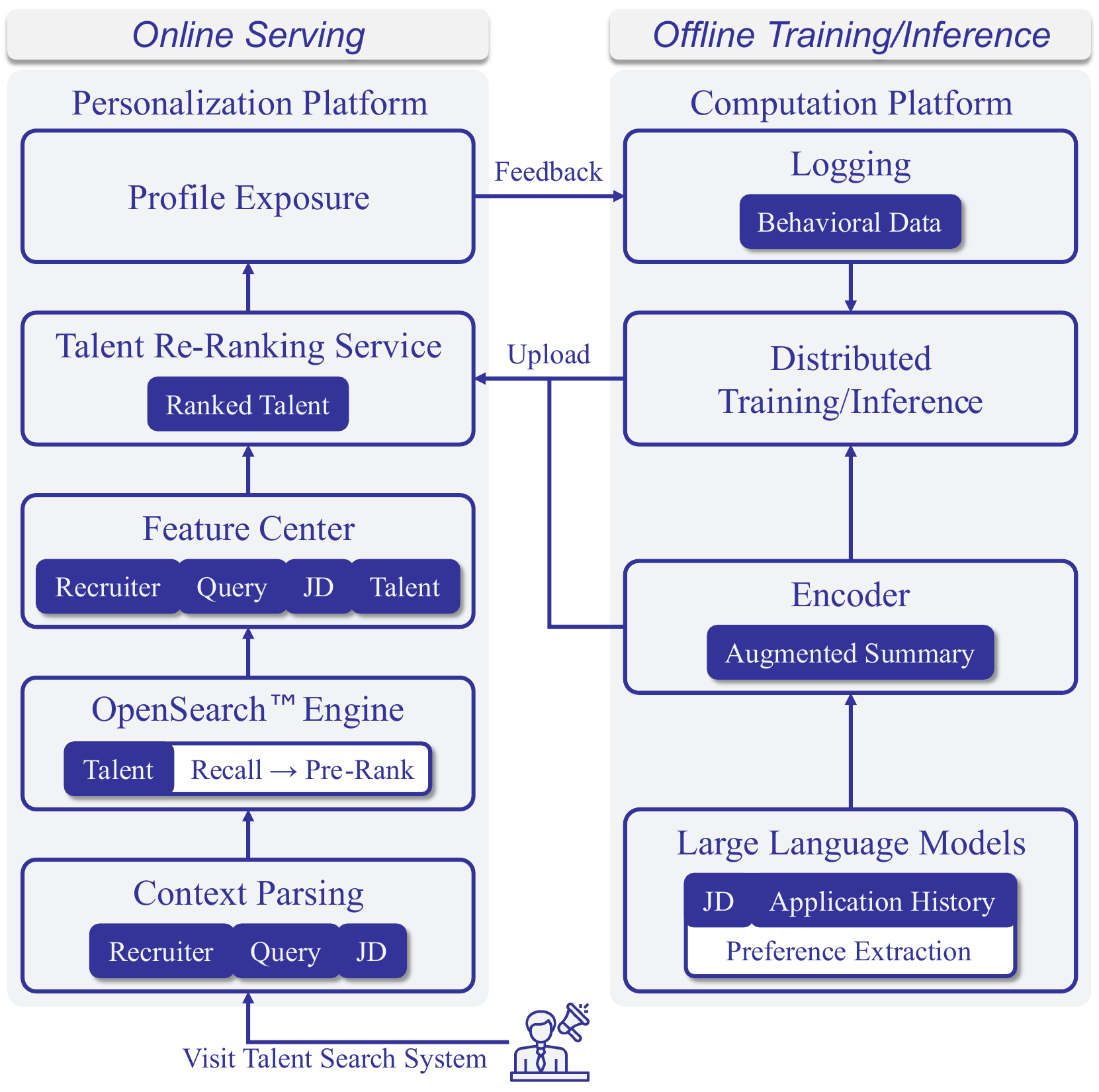}
  \caption{Workflow of our talent search system.}
  \label{fig:workflow}
\end{figure}

As illustrated in \Cref{fig:workflow}, our talent search system consists of two
main components: online serving and offline training/inference. The online
serving side handles real-time recruiter queries by parsing context, retrieving
candidate profiles through the OpenSearch\textsuperscript{\texttrademark}
engine~\cite{opensearch}, and re-ranking results using the proposed Talent
Re-Ranking Service. The offline module supports training and inference, where
behavioral data is logged, processed, and used to train models on a distributed
platform. Our framework enhances this pipeline by integrating LLMs for
recruitment preference extraction and generating augmented summaries, which are
then encoded by the proposed Encoder. These representations are uploaded to the
online system for re-ranking, enabling personalized and context-aware talent
recommendations.

\section{LLM Prompting}
\label{sec:prompt}
% We present an example of our prompt used for the LLM, as shown in
% Figure~\ref{fig:prompt-1} and Figure~\ref{fig:prompt-2}. We first ask the LLM
% to act as a seasoned HR expert and introduce the task, and then we list some
% requirements and constraints, and the LLM is required to perform the analysis
% step by step. We also add an output format example and the inputs in the
% prompt. Through meticulous prompt design, we can effectively mitigate the
% hallucination issues in large language models.

% We provide an example of the prompt used for the LLM, illustrated in
% \Cref{fig:prompt-1} and \Cref{fig:prompt-2}. The prompt instructs the LLM to
% assume the role of an experienced HR specialist and introduces the task
% context. It then specifies a set of requirements and constraints, guiding the
% LLM to perform the analysis in a step-by-step manner. To further ensure
% consistency and structure in the generated output, we include an explicit
% output format template along with example inputs. This carefully crafted prompt
% design helps reduce hallucination and improves the reliability of the
% LLM-generated recruitment preference summaries.

\begin{figure}[h]
    \centering
    \includegraphics[width=\columnwidth]{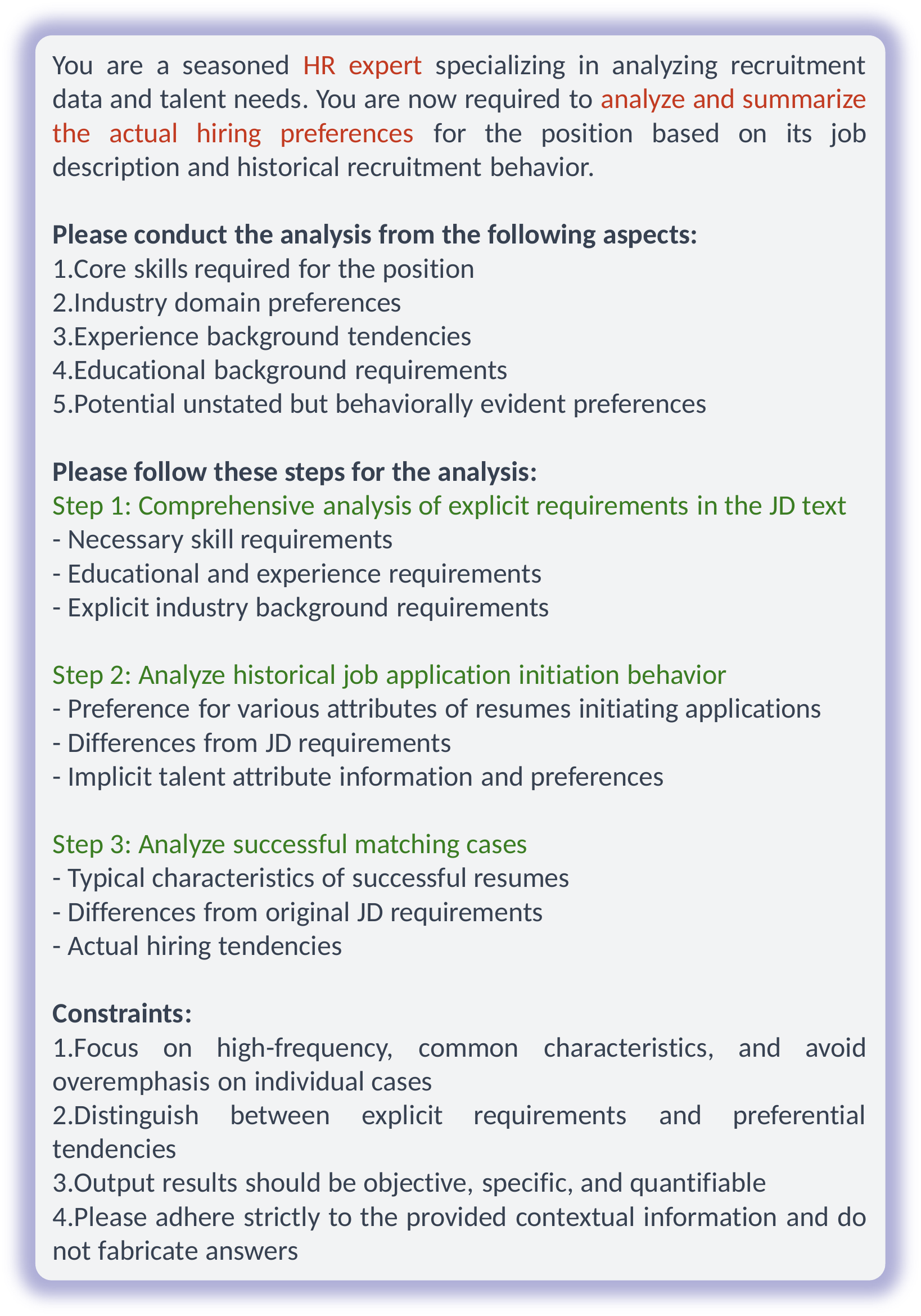}
    \caption{LLM Prompt: System Instruction}
    \label{fig:prompt-1}
\end{figure}

\begin{figure}[!h]
    \centering
    \includegraphics[width=\columnwidth]{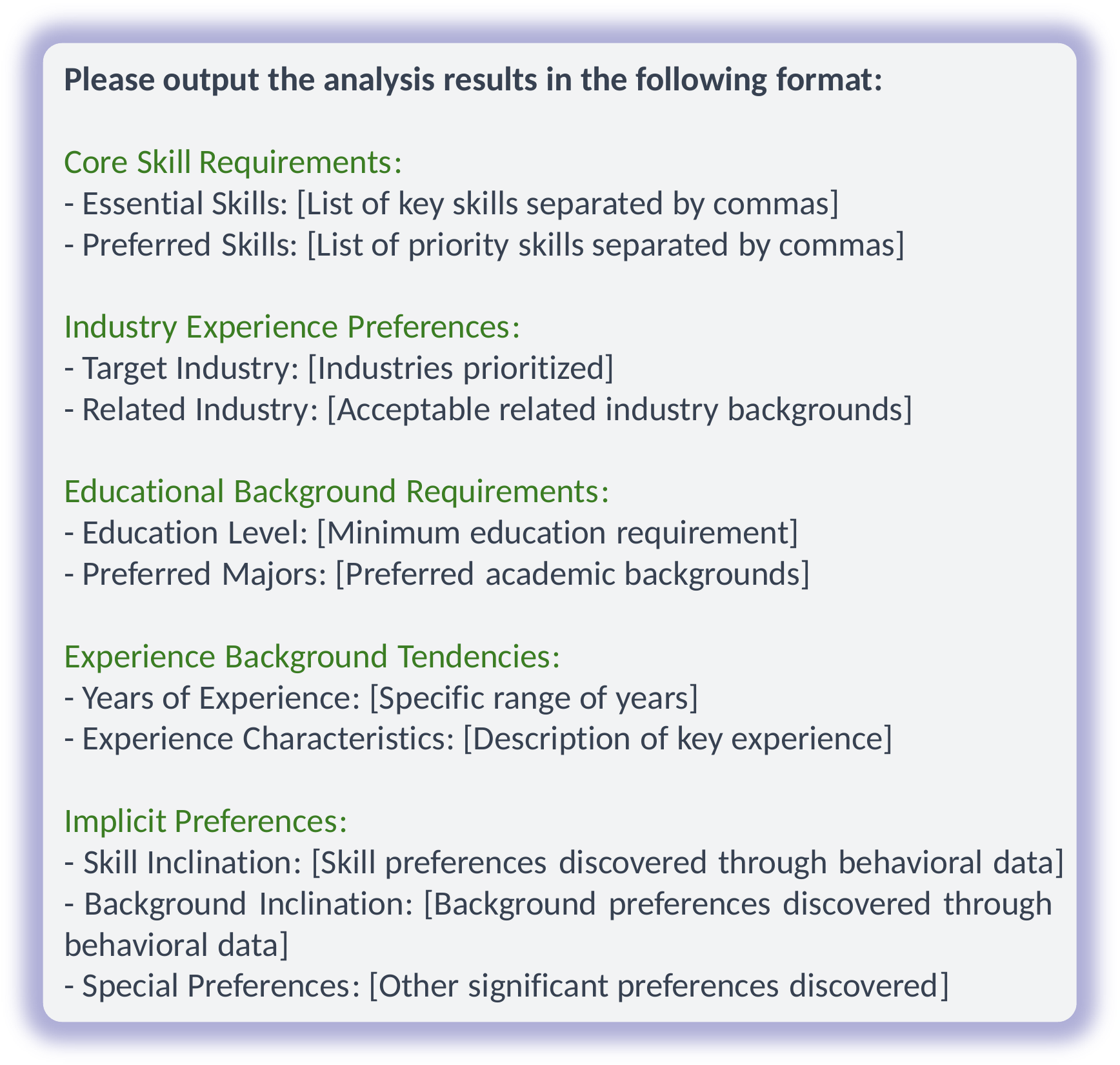}
    \caption{LLM Prompt: Output}
    \label{fig:prompt-2}
\end{figure}

We illustrate the prompt used for the LLM in \Cref{fig:prompt-1} and
\Cref{fig:prompt-2}. The prompt frames the LLM as an experienced HR specialist,
introduces the task context, and outlines specific requirements and constraints
to guide step-by-step analysis. To ensure consistency and structure, we also
include a predefined output format and example inputs. This prompt design helps
reduce hallucinations and enhances the quality of the generated recruitment
preference summaries.

% \section{Multi-task Learning Models for Recommendation}
\section{Multi-task Learning Models}
\label{sec:appendix_related_work}
We review recent multi-task learning (MTL) models in recommendation systems,
focusing on those built upon the widely adopted Embedding-Tower
architecture~\cite{huang2013learning}, where inputs are embedded and passed
through task-specific towers.

ESSM~\cite{ma2018entire} enhances CVR prediction by introducing auxiliary CTR
and CTCVR tasks and sharing embedding parameters between CVR and CTR.
\citet{chen2019co} employ hierarchical multi-pointer
co-attention~\cite{tay2018multi} to model task correlations, improving
performance in both recommendation and explanation tasks.

MMoE~\cite{ma2018modeling}, adopted in large-scale systems such as YouTube's
video recommender~\cite{zhao2019recommending}, introduces a set of shared
expert networks whose outputs are routed via task-specific gating mechanisms.
This design allows tasks to selectively leverage shared knowledge while
preserving their unique modeling needs.

% Despite these advances, MTL often faces challenges such as negative transfer
% and the seesaw effect, where optimizing one task harms another. To mitigate
% these issues, PLE~\cite{tang2020progressive} separates shared and task-specific
% expert towers and introduces progressive routing, effectively managing task
% interference and enabling better disentanglement of shared versus private
% knowledge. Most recently, STEM~\cite{su2024stem} proposes a novel paradigm that
% unifies shared and task-specific embeddings under an All-Forward, Task-Specific
% Backward (AF-TB) gating mechanism. This design explicitly enhances
% task-specific representations while allowing effective knowledge sharing,
% yielding strong empirical performance across multi-task recommendation
% benchmarks.
Despite their success, MTL often face challenges such as negative transfer and
the seesaw effect, where improving one task may degrade another. To address
this, PLE~\cite{tang2020progressive} separates shared and task-specific experts
and introduces progressive routing to better manage task interference and
disentangle shared from private knowledge. More recently,
STEM~\cite{su2024stem} presents a unified embedding paradigm with an
All-Forward, Task-Specific Backward gating mechanism, which strengthens
task-specific representations while supporting knowledge sharing, achieving
strong results across multi-task recommendation benchmarks.

\section{Implementation Details}
\label{sec:implementation}
% Here, we introduce the key implementation details in our experiments. The
% dimensions of the user, talent and query ID embeddings are all 32, while the
% dimensions of the query and job description text embeddings are both 1024.
% Dimension of $c_0$ is 2048, and the hidden sizes of $W^c$ and $W^g$ are both
% 2048. The number of attention heads $h$ in multi-head attention is 1. For the
% MMoE related hyperparameters, the number of experts $n^e$ is 3, the hidden
% sizes of $f_1^g$, $f_2^g$, $f_1^e$, $f_2^e$ and $f_3^e$ are 32, 16, 128, 64
% and 32, respectively. For the multi-task learning module, $\mathbf{DNN}_1$
% contains three layers whose hidden sizes are 512, 256 and 128 respectively;
% $\mathbf{DNN}_2$ also contains three layers whose hidden sizes are 256, 128
% and 64 respectively. The weights of the three losses $\lambda_1$,
% $\lambda_2$, $\lambda_3$ are equal and are all set to 1.0. For model
% training, we use a batch size of 1024 and a learning rate of 1e-5. The model
% is trained for nearly 20000 steps with a dropout rate of 0.2.

% For the baselines, we tune the hyperparameters for them to achieve their
% optimal performance on our dataset.

This section outlines the key implementation details of our experiments.

\paragraph{Embedding Dimensions} The ID embeddings for recruiters, queries, and
talents are all set to 32 dimensions. The query and job description text
embeddings are both 1024-dimensional. The concatenated vector $c_{0}$ has a
dimensionality of 2048, and the hidden sizes of $W^{(c)}$ and $W^{(g)}$ are
both 2048. For the multi-head attention module, the number of attention heads
is set to 1.

\paragraph{MMoE Configuration} The number of experts in the MMoE module is 3.
The hidden sizes are 128, 64 and 32 for the expert networks, and 32 and 16 for
the gating networks.

\paragraph{Multi-task Learning Module} We use two DNN towers for task-specific
modeling: (i) the first tower has three layers with hidden sizes of 512, 256,
and 128, and (ii) the second tower has three layers with hidden sizes of 256,
128, and 64. All loss weights $\lambda_1, \lambda_2, \lambda_3$ are set to 1.0.

\paragraph{Training Setup} The model is trained with a batch size of 1024,
learning rate of $1 \times 10^{-5}$, dropout rate of 0.2, and for approximately
20{,}000 steps.

\paragraph{Baseline Configuration} For all baseline models, we perform
hyperparameter tuning to ensure each achieves its optimal performance on our
dataset.

% We summarize the key implementation settings used in our experiments.

% \paragraph{Embedding Dimensions} Recruiter, query, and talent ID embeddings are
% all 32-dimensional. Query and JD text embeddings are 1024-dimensional. The
% concatenated vector $c_0$ has dimension 2048, with $W^{(c)}$'s and $W^{(g)}$'s
% hidden sizes also set to 2048. The multi-head attention module uses a single
% head.
%
% \paragraph{MMoE Configuration} The MMoE module uses 3 experts. Expert hidden
% sizes are 128, 64, and 32; gating networks use hidden sizes of 32 and 16.
%
% \paragraph{Multi-task Learning Module} We employ two DNN towers for
% task-specific learning. The first has hidden sizes of 512, 256, and 128; the
% second has 256, 128, and 64. All task loss weights $\lambda_1$, $\lambda_2$,
% and $\lambda_3$ are set to 1.0.
%
% \paragraph{Training Setup} The model is trained with a batch size of 1024,
% learning rate of $1 \times 10^{-5}$, dropout rate of 0.2, and for approximately
% 20{,}000 steps.

% \paragraph{Baselines} All baseline models undergo hyperparameter tuning to
% ensure competitive performance on our dataset.

% \section{Case Study}
% \label{app:case_study}
% \Cref{fig:case_study} shows the job description and candidate rankings for the
% ``Search Product Manager'' case study referenced in the main text.

\end{document}